\documentclass[useAMS,usenatbib,a4paper,preprint]{mn2e}
\usepackage{graphicx}
\usepackage{amsmath}
\usepackage{txfonts}
\usepackage{mathrsfs}

\usepackage{color}

\def\aap{A\&A}
\def\apjl{ApJ}

\def\apjs{ApJS}

\def\apjs{ApJ Supplements}

\def\apjl{ApJ Letters}

\def\aap{A\&A}

\newcommand{\be}{\begin{equation}}
\newcommand{\ee}{\end{equation}}
\newcommand{\bea}{\begin{eqnarray}}
\newcommand{\eea}{\end{eqnarray}}

\title[The Scale of Homogeneity]{The Scale of Homogeneity in the $R_{\rm h}=ct$ Universe}
\author[Fulvio Melia]{Fulvio Melia\thanks{John Woodruff Simpson
Fellow. E-mail: fmelia@email.arizona.edu}\\
Department of Physics, The Applied Math Program, and Department of Astronomy,
The University of Arizona, AZ 85721, USA}

\begin{document}
\date{} 
\pagerange{\pageref{firstpage}--\pageref{lastpage}} 
\pubyear{2023}

\maketitle

\label{firstpage}

\begin{abstract} Studies of the Universe's transition to smoothness in the context 
of $\Lambda$CDM have all pointed to a transition radius no larger than $\sim 300$ Mpc. 
These are based on a broad array of tracers for the matter power spectrum, including 
galaxies, clusters, quasars, the Ly-$\alpha$ forest and anisotropies in the cosmic 
microwave background. It is therefore surprising, if not anomalous, to find many 
structures extending out over scales as large as $\sim 2$ Gpc, roughly an order of 
magnitude greater than expected. Such a disparity suggests that new physics may be 
contributing to the formation of large-scale structure, warranting a consideration 
of the alternative FLRW cosmology known as the $R_{\rm h}=ct$ universe. This model
has successfully eliminated many other problems in $\Lambda$CDM. In this paper, 
we calculate the fractal (or Hausdorff) dimension in this cosmology as a function 
of distance, showing a transition to smoothness at $\sim 2.2$ Gpc, fully 
accommodating all of the giant structures seen thus far. This outcome adds 
further observational support for $R_{\rm h}=ct$ over the standard model. 
\end{abstract}


\section{Introduction}\label{intro}
The standard model of cosmology ($\Lambda$CDM; \citealt{Ostriker:1995}) 
is based on the Friedmann-Lema\^itre-Robertson-Walker (FLRW) spacetime, 
whose metric coefficients reflect the symmetries assumed by the cosmological
principle, i.e., isotropy and homogeneity \citep{Melia:2020}, at least
on large scales. Of the two, isotropy is easier to measure, since the 
observations can be carried out from a single vantage point, using the 
cosmic microwave background (CMB; \citealt{Schwarz:2016,Planck:2016}), 
galaxy and radio source counts \citep{Rameez:2018}, weak lensing convergence
\citep{Marques:2018}, and Type Ia supernovae \citep{Andrade:2018}.

Homogeneity is measured less reliably because we can only probe structure
down our past lightcone, not directly on a time-constant hypersurface
\citep{Clarkson:2012}. One must interpret these measurements in the context
of how well the model predictions are confirmed by the observed structure
evolution within the past lightcone. Observational tests of homogeneity
have been carried out by several workers, using various tracers of the matter
distribution, including galaxies, clusters of galaxies, the Ly-$\alpha$
forest and quasars 
\citep{Hogg:2005,Scrimgeour:2012,Alonso:2015,Laurent:2016,Ntelis:2017,Goncalves:2018,Secrest:2021,Camacho:2022}.
These studies typically employ the so-called fractal (or correlation)
dimension (see Eq.~\ref{eq:H} below) to characterize the spatial scale
at which homogeneity is attained. 

In the majority of cases, the transition is quoted at $\sim 100-200\;
{\rm Mpc}/h$, where $h\equiv H_0/100$ km s$^{-1}$ Mpc$^{-1}$ is the 
scaled Hubble parameter. Throughout this paper, we shall conveniently 
adopt the {\it Planck} value optimized for $\Lambda$CDM, i.e., 
$h=0.6732$ \citep{Planck:2016}. Some studies possibly stretch the range 
to $\sim 260\;{\rm Mpc}/h$ \citep{Yadav:2010}. Very importantly, however, 
no study has yet shown a termination of the clustering on scales larger 
than $\sim 300$ Mpc. 

So it comes as a surprise to learn that cosmic structures can form 
bigger---even much bigger---than this limit. A sample discovered thus 
far is shown in Table~\ref{tab1} below. This includes the Giant GRB
Ring, traced by 9 gamma-ray bursts (GRBs) at $z\sim 0.82$, with a 
diameter of $\sim 1720$ Mpc \citep{Balazs:2015}. With a probability of 
$\sim 2\times 10^{-6}$ of this being merely the result of random
GRB count rate fluctuations, the structure appears to be seen as
a projection of a shell on the plane of the sky. The largest structure
seen thus far appears to be the Hercules-Corona Borealis Great Wall
(HCB Great Wall), also identified via a large GRB cluster at $z\sim 2$ 
with a size of $2000-3000$ Mpc \citep{Horvath:2014}. Its angular
size remarkably covers one-eighth of the sky. Other large structures 
have been identified via quasar associations, e.g., the Huge-LQG 
(Huge-Large Quasar Group) centered at $z=1.27$ with a size of 
$1240$ Mpc \citep{Clowes:2012,Hutsemekers:2014}, and in galaxy 
surveys, including the Sloan Great Wall of galaxies at 
$z\sim 0.073$ with a length of $450$ Mpc \citep{Gott:2005}.

\begin{table*}
\begin{center}
\begin{minipage}{580pt}
\caption{Sample of very large structures identified thus far.}\label{tab1}%
{\footnotesize
\begin{tabular}{@{}llll@{}}
\hline\hline \\
Name & Mean redshift & Proper Size & Reference \\
     &        & (Mpc)                  & \\ \\
\hline \\
HCD Great Wall & $\sim 2$ & 2000-3000 & \cite{Horvath:2014,Christian:2020} \\
Giant GRB Ring & $0.82$ & 1720 & \cite{Balazs:2015} \\
Correlated LQG Orientations & $1.0-1.8$ & 1600 & \cite{Friday:2022} \\
U1.27, Huge-LQG & $1.27$ & 1240 & \cite{Clowes:2012,Hutsemekers:2014} \\
Giant Arc & $\sim 0.8$ & 1000 & \cite{Lopez:2022} \\
Coherent Quasar Polarizations & $1-2$ & 1000 & \cite{Hutsemekers:2005} \\
U1.11 & $1.11$ & 780 & \cite{Clowes:2012} \\
U1.28, CCLQG & $1.28$ & 630 & \cite{Clowes:2012} \\
Sloan Great Wall & $0.073$ & 450 & \cite{Gott:2005} \\
South Pole Wall & $0.04$ & 420 & \cite{Pomarede:2020} \\
Blazar LSS & $\sim 0.35$ & 350 & \cite{Marcha:2021} \\
Local Void & $<0.07$ & 300 & \cite{Whitbourn:2016} \\ \\
\hline
\end{tabular}
}
\end{minipage}
\end{center}
\end{table*}

But there are other observational indications that the 
generic assumption of smoothness on scales exceeding $\sim 200$ Mpc 
in the standard model is not well supported by the data. For example,
in their statistical analysis of the visible matter distribution,
\cite{Coleman:1992} concluded that it is fractal and multifractal
on such scales, without any evidence of termination, i.e.,
homogenization. Subsequent work \citep{Labini:1998,Labini:2009} 
confirmed that galaxy structures are irregular and self-similar, 
consistent with fractal correlations up to scales exceeding $\sim 1$ 
Gpc. The implication is that the distribution of matter is not 
analytic and cannot be described in terms of a simple average 
density.  These inhomogeneities therefore appear to challenge
our conventional view of cosmology \citep{Labini:2011}. It has
even been suggested that an otherwise smooth metric, such as 
the Lema\^itre-Tolman-Bondi universe, that is able to describe,
on average, a fractal distribution of matter, may explain
cosmic acceleration as a purely fractal phenomenon
\citep{Cosmai:2019}. As we shall see in this paper, however, 
the transition to smoothness is heavily model dependent so, 
while this type of statistical analysis does not comport very 
well with the small transition to smoothness scale predicted 
by $\Lambda$CDM, it would be fully consistent with an alternative 
model that predicts a much larger scale of homogeneity, well 
beyond the fractal domain. 

The discovery of super large structures comes with several 
important caveats, that we shall discuss in greater detail below. 
First, we are actually dealing with two overlapping issues: the 
{\it average} transition to smoothness versus statistical fluctuations 
forming anomalously large structures. These are not necessarily incompatible
with each other \citep{Nadathur:2013}. Second, it is not certain
that we have detected the largest structures yet. In their implementation
of the Zipf-Mandelbrot law to the size distribution of superclusters
of galaxies, \cite{DeMarzo:2021} conclude that none of the currently
available catalogs are sufficiently large for us to have seen
a truncation in their extent. Deeper redshift surveys may yet
uncover even larger structures than the greatest already observed.

No matter how one interprets these results, however, the discovery
of structures an order of magnitude larger than the transition to
smoothness begs for an explanation. In this paper, we examine this
issue in the context of the alternative FLRW cosmology known as the 
$R_{\rm h}=ct$ universe \citep{Melia:2007,MeliaShevchuk:2012,Melia:2020}. 
This model has been discussed extensively in both the primary and 
secondary literature, so we won't dwell on its foundational and 
observational details in this paper, except to highlight the fact 
that, viewed as an advanced version of $\Lambda$CDM, it actually 
solves many of the current standard model's conflicts and tensions 
\citep{Melia:2022e}. The most recent comparative study, based on 
the surprising detection by {\it JWST} of large well formed galaxies 
at much higher redshifts than can be accommodated by $\Lambda$CDM
\citep{Pontoppidan:2022,Finkelstein:2022,Treu:2022,Robertson:2022}, 
has shown that the timeline in $R_{\rm h}=ct$ is much more compatible 
with the implied formation of structure in the early Universe 
\citep{Melia:2023b}. 

The improvements introduced by this `advanced' version of the standard
model are therefore not merely cosmetic. It solves most, if not all,
of $\Lambda$CDM's long-standing problems. The consistency between 
the formation of structure in $R_{\rm h}=ct$ and the most recent 
observations by {\it JWST} therefore suggests that the creation of 
the anomalously large features listed in Table~\ref{tab1} may be 
more compatible with this alternative FLRW cosmology.

Given that the creation of structure is tightly related to the physics 
of star, galaxy, and cluster formation, the anomalies we explore in 
this paper, specifically how $\sim 2$ Gpc structures could possibly 
have formed in an FLRW Universe, are of significant importance to 
ongoing and future surveys of the Universe at intermediate and 
high redshifts.

\section{Scale of Homogeneity}\label{homo}
A technique commonly used to determine the spatial distance at which the
large-scale structure transitions to homogeneity is based on multifractal 
analysis, in which the fractal dimension, $D_2$, of the underlying point 
distribution approaches the ambient dimension, $D=3$, of the space where 
the points lie \citep{Bagla:2008,Yadav:2010}. In reality, $D_2$ never 
quite reaches 3, so an alternative, less definitive, criterion needs to 
be defined. An example of this was the suggestion by \cite{Yadav:2010}
to use the distance at which the deviation, $(\Delta D_2)_{\rm clus}$, 
of the fractal dimension due to clustering becomes smaller than its 
statistical dispersion. 

The definition of a `homogeneity scale' tends to be somewhat arbitrary
because the approach to homogeneity is gradual, not abrupt. The proposal 
by \cite{Yadav:2010} has been questioned because in deriving the 
dispersion they ignored the contribution to variance from the survey 
geometry and selection function. The dispersion calculated from a real 
survey is therefore not easily interpreted using this definition, 
certainly not when comparing different samples derived with inconsistent 
selection criteria. A more commonly used definition of the `homogeneity 
scale' today is the distance, $r_{\mathcal{H}}$, at which the deviation 
of the fractal dimension due to clustering drops to $1\%$ of $D$ 
\citep{Scrimgeour:2012}.

Detailed information on clustering within the distribution is inferred
from a correlation integral,
\begin{equation}
C_2\equiv {1\over NM}\sum_{i=1}^M n_i(r)\;,\label{eq:corr}
\end{equation}
where $N$ is the number of points in the distribution, $M$ is the
number of spheres of radius $r$ (much smaller than the overall
size of the space), and $n(r)$ is the number of points within a
distance $r$ from the $i^{\rm th}$ point:
\begin{equation}
n_i(r)\equiv \sum_{j=1}^N\Theta(r-|s_i-s_j|)\;.\label{eq:ni}
\end{equation}
Here, $\Theta$ is the Heaviside function, and $|s_i-s_j|$ is the
distance between each pair of points $(i,j)$.

Depending on how the points are clustered (if at all), the fractal
dimension changes with $r$---and presumably approaches that of the
space ($D$) asymptotically. The Minkowski-Bouligand dimension is 
defined as
\citep{Yadav:2010}
\begin{equation}
D_2(r) \equiv {d\,\log C_2(r)\over d\,\log r}\;.\label{eq:D2}
\end{equation}
If the distribution is homogeneous, then clearly $C_2(r)\propto
r^3$ and $D_2(r) = 3$, for a space with dimension $D=3$.

The geometrical sampling of a given survey is not perfect, however,
so a scaled correlation integral was introduced \citep{Scrimgeour:2012}, 
which is just $C_2(r)$ normalized by the same quantity in a random 
homogeneous distribution (including selection effects, if present):
\begin{equation}
{\mathcal{N}}\equiv {C_2(r)\over \bar{C}_2(r)}\;,\label{eq:mathcalN}
\end{equation}
where 
\begin{equation}
\bar{C}_2(r)\equiv {4\pi\over 3}r^3\bar{n}\;,\label{eq:barC}
\end{equation}
in terms of the average number density, $\bar{n}$.

Equation~(\ref{eq:corr}) may be written in terms of the
two-point correlation function, $\xi(r)$, starting
with the differential \citep{Ntelis:2017}
\begin{equation}
dC_2(r) = \bar{n}\left[1+\xi(r)\right]dV\;,\label{eq:C2xi}
\end{equation}
where $dV$ is the differential volume around a given particle. The
integrated correlation integral is then simply
\begin{equation}
C_2(r) = 4\pi\bar{n}\int_0^r\left[1+\xi(u)\right]u^2\,du\;,\label{eq:C2int}
\end{equation}
under the assumption of isotropy, for which $\xi({\bf r})=\xi(r)$. It is
straightforward to see that Equation~(\ref{eq:mathcalN}) then becomes
\begin{equation}
{\mathcal{N}}(r) = 1+{3\over r^3}\int_0^r\xi(u)u^2\,du\;.\label{eq:mathcalNxi}
\end{equation}

It is not difficult to see from Equations~(\ref{eq:D2}) and (\ref{eq:mathcalN})
that the fractal correlation dimension may also be written as
\begin{equation}
D_2(r) = {d\,\log {\mathcal{N}}\over d\,\log r}+D\;.\label{eq:D2mathcalN}
\end{equation}
We may thus define the homogeneity index, ${\mathcal{H}}$, a fractal or
Hausdorff dimension, which in this context represents the deviation
$(\Delta D_2)_{\rm clus}$, as
\begin{equation}
{\mathcal{H}}\equiv {d\,\log {\mathcal{N}}\over d\,\log r}.\label{eq:H}
\end{equation}
Complete homogeneity is thus reached when ${\mathcal{H}}=0$ (or $D_2=D$).

If we now define the volume-averaged two-point correlation function
\begin{equation}
\bar{\xi}(r) \equiv {3\over r^3}\int_0^r \xi(u)u^2\,du\;,\label{eq:barxi}
\end{equation}
we see that 
\begin{equation}
{\mathcal{N}}=1+\bar{\xi}\;,\label{eq:Nbarxi}
\end{equation}
and therefore
\begin{equation}
{\mathcal{H}}={D\over 1+\bar{\xi}}\left[\xi-\bar{\xi}\right]\;.\label{eq:Hfinal}
\end{equation}
Note that this expression is not restricted by any assumption concerning
the magnitude of $\xi$ or $\bar{\xi}$. For the rest of this paper, we shall
follow current convention and assume that homogeneity is reached when
${\mathcal{H}}$ is $<1\%$ of $D$, i.e., for $|{\mathcal{H}}|<0.03$ in
3-dimensional space \citep{Scrimgeour:2012,Laurent:2016,Ntelis:2017,Goncalves:2018}.

\section{The two-point correlation function}\label{two-point}
The two-point correlation function may be expressed in terms of its Fourier
transform, the power spectrum, $P(k)$, according to
\begin{equation}
\xi(s) = {1\over (2\pi)^{3/2}}\int P(k) e^{-i{\bf k}\cdot{\bf s}}d^3k\;.\label{eq:xi}
\end{equation}
For an isotropic distribution, one may integrate this expression over angles and obtain 
the one-dimensional integral
\begin{equation}
\xi(s)={1\over 2\pi^2}\int_0^\infty k^2\,dk\,P(k) j_0(ks)\;,\label{eq:xi2}
\end{equation}
where 
\begin{equation}
j_0(x)={\sin(x)\over x}\label{eq:j0}
\end{equation}
is a spherical Bessel function.

As discussed above, the matter distribution may be measured using several different 
kinds of tracers, including galaxies, quasars, clusters of galaxies, fluctuations in 
the CMB and the Ly-$\alpha$ forest. Previous work has shown that all of these 
measurements tend to yield similar values of the transition radius, typically in 
the range $\sim 100-200\; {\rm Mpc}/h$. A principal concern of our comparative
analysis is the proper recalibration of these data for each individual cosmology,
given that densities and distances vary from one model to the next. Since we
are primarily interested in finding variations from $\sim 300$ Mpc to
$\sim 2000$ Mpc, rather than incremental differences from one type of object
to the next within the same background cosmology, we shall conveniently choose 
a set of data for which the recalibration is quite straightforward, i.e., 
we shall simply use the fluctuations in the CMB and the Ly-$\alpha$ forest, 
more fully described in \cite{Yennapureddy:2021}.

\begin{figure}
\centering
\includegraphics[scale=0.5]{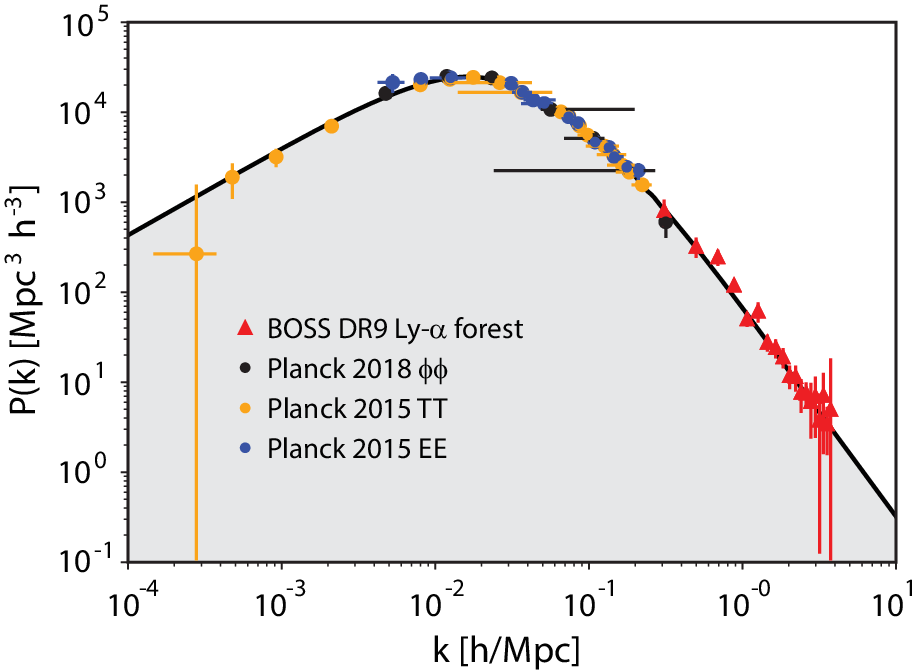}
\caption{The matter power spectrum based on observations of the
CMB (orange, blue and black dots), and the Ly-$\alpha$ survey (red
triangles), compared with the power spectrum predicted by $\Lambda$CDM 
(solid black curve). Here, $h$ is the parameter $H_0/(100\;{\rm km}\;{\rm s}^{-1}
\;{\rm Mpc}^{-1})$. Throughout this paper, we assume the value $h=0.6732$,
consistent with the latest {\it Planck} measurements (Planck Collaboration
et al. 2016). (Adapted from Yennapureddy and Melia 2021)}
\label{fig1}
\end{figure}

\begin{figure}
\centering
\includegraphics[scale=0.5]{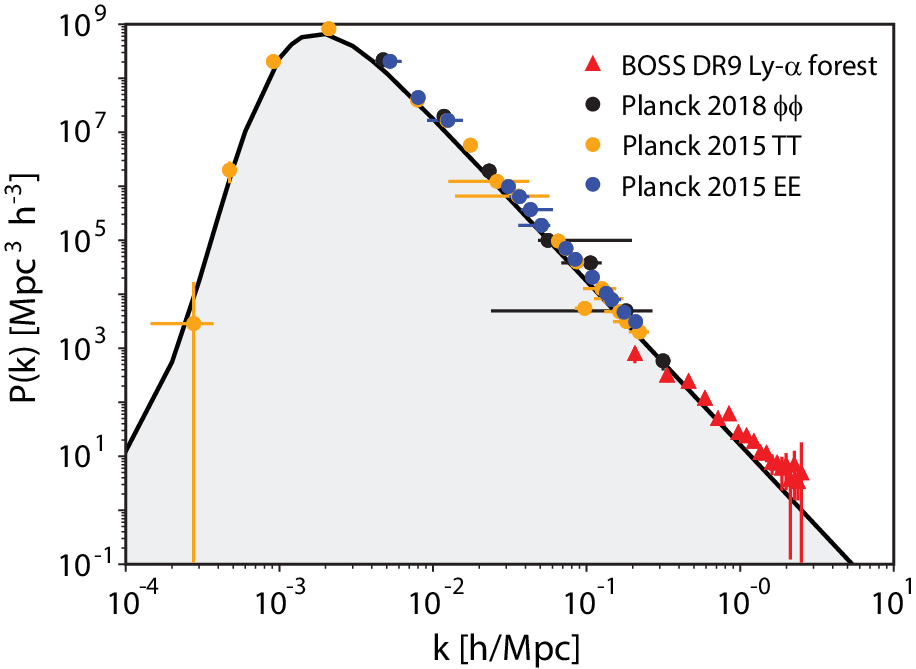}
\caption{Same as Fig.~1, except here the data are calibrated for (and
compared with) the $R_{\rm h}=ct$ cosmology (solid black curve). As 
before, $h=0.6732$ is the parameter $H_0/(100\;{\rm km}\;{\rm s}^{-1}
\;{\rm Mpc}^{-1})$, consistent with the latest {\it Planck} measurements
(Planck Collaboration et al. 2016). (Adapted from Yennapureddy and
Melia 2021)}
\label{fig2}
\end{figure}

These data are compared to the theoretical predictions in Figures~(\ref{fig1})
and (\ref{fig2}) for $\Lambda$CDM and $R_{\rm h}=ct$, respectively, calculated 
from the assumed primordial power spectrum and the transfer functions in these 
models.  These two figures look quite different from each other because the
data are not model-independent. The CMB measurements in these diagrams are 
shown as blue, orange and black circular dots. These are calculated from the 
CMB observations \citep{Planck:2016} using the approach of \cite{Tegmark:2002}. 

The Ly-$\alpha$ forest (shown as red triangles in Figs.~\ref{fig1}
and \ref{fig2}) is due to the absorption along the line-of-sight of 
high-redshift quasar spectra. The fluctuations result from the 
inhomogeneous neutral hydrogen within the photo-ionized intergalactic 
medium. Since the underlying mass density is related to the optical 
depth of the Ly-$\alpha$ absorption line, the Ly-$\alpha$ forest is a 
proxy for the matter power spectrum. The amplitude of the fluctuations 
is itself model-dependent, however, so the Ly-$\alpha$ data must also be 
recalibrated for each individual cosmology. A caveat here is that the 
hydro simulations used to translate the absorption profile into the 
matter fluctuations are based on the behavior of dark matter only, and 
may not have included all of the relevant physics. Uncertainties in the 
reionization history, the ionizing background and its fluctuations may 
therefore be propagating in unreliable ways through the reconstruction 
of $P(k)$. As such, the Ly-$\alpha$ data may not be as reliable as the 
other points shown in this figure (see \citealt{Yennapureddy:2021} for 
all the details).

Insofar as this paper is concerned, the most important conclusion we can 
draw from Figures~\ref{fig1} and \ref{fig2} is the remarkable consistency 
between the predicted power spectrum in {\it both} models and all of the 
available data. We can see the broad agreement between theory 
and the observations quantitatively via the reduced $\chi^2$ of the fits. 
The theoretical curves shown here are calculated simply using the 
{\it Planck} parameters without any optimization. For $\Lambda$CDM, one 
finds $\chi^2_{\rm dof} \approx 1.31$; the corresponding value for 
$R_{\rm h}=ct$ is $\chi^2_{\rm dof}\approx 1.3$. The fits could be 
improved slightly with additional optimization, but that is not really 
the focus of this paper. Demonstrating this consistency is the main 
reason for showing both the theoretical curves and the data, because we 
are thus confident that using the theoretically calculated curve for 
$P(k)$ provides an accurate assessment of the two-point correlation 
function predicted in both $\Lambda$CDM and $R_{\rm h}=ct$ via 
Equation~(\ref{eq:xi2}). In other words, our estimation of the 
transition radius is fully based on the `observed' two-point correlation
function in both models. {\sl It is not merely an untested theoretical
prediction.}

\begin{figure*}
\centering
\includegraphics[scale=0.9]{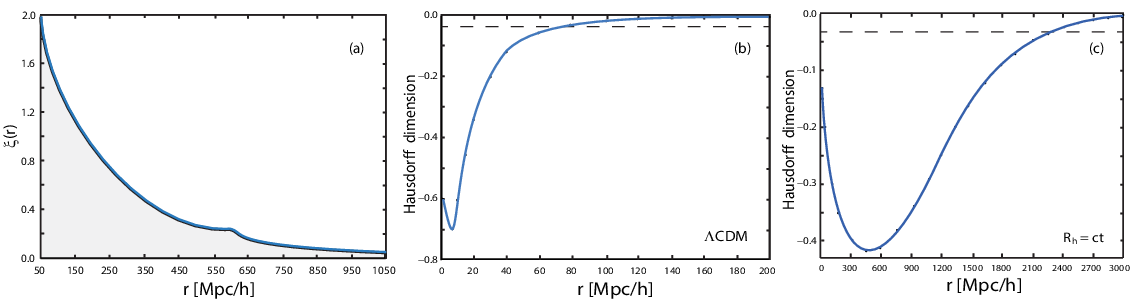}
\caption{(a)~The two-point correlation function, $\xi(r)$, 
as a function of $r/h$, calculated in the $R_{\rm h}=ct$ universe from the 
matter power spectrum in Fig.~\ref{fig2}. (b)~Homogeneity index estimated 
from the matter power spectrum in Fig.~1, calibrated for the concordance 
$\Lambda$CDM model. Using the nominal definition of a threshold to 
smoothness when {$|\mathcal{H}|$} is less than $1\%$ of the space 
dimension, $D=3$, we infer a comoving transition distance 
$r_{\mathcal{H}}^{\Lambda{\rm CDM}} \sim 80\;{\rm Mpc}/h$.
(c)~Homogeneity index estimated from the matter power spectrum
in Fig.~2, calibrated for the $R_{\rm h}=ct$ universe. Using the
smoothness transition at {$|\mathcal{H}|$} equal to $1\%$ of the
space dimension, $D=3$, we infer a comoving transition distance
$r_{\mathcal{H}}^{R_{\rm h}=ct} \sim 2,290\;{\rm Mpc}/h$.}
\label{fig3}
\end{figure*}

The two-point correlation function derived from Figure~(\ref{fig1}) 
for $\Lambda$CDM is well known, so we don't need to reproduce it here. 
The corresponding result for $R_{\rm h}=ct$, derived from 
Figure~(\ref{fig2}), is shown in Figure~(\ref{fig3}a).

\section{Scale of homogeneity in $R_{\rm h}=ct$}\label{scale}
As a sanity check, let us first calculate the fractal (or Hausdorff)
dimension for $\Lambda$CDM to demonstrate consistency with previous
work. ${\mathcal{H}}(r)$ in this model is shown as a function of $r$
in Figure~(\ref{fig3}b). The conventional definition of a threshold
to smoothness, i.e., $|{\mathcal{H}}|<0.03$, implies that the
$\Lambda$CDM universe becomes homogeneous on scales larger
than $\sim 80\;{\rm Mpc}/h$. This outcome is entirely consistent
with other estimations based on the use of various tracers of the
matter distribution. ${\mathcal{H}}(r)$ is effectively zero by
the time $r$ gets to several hundred Mpc and above. This is the
reason, of course, why the much larger structures listed in 
Table~\ref{tab1} appear to be anomalous in this model. 

The corresponding fractal (or Hausdorff) dimension for the $R_{\rm h}=ct$
universe is shown as a function of $r$ in Figure~(\ref{fig3}c). Evidently,
homogeneity in this model is reached on scales greater than about
$2200\;{\rm Mpc}/h$, about an order of magnitude larger than its
counterpart in the standard model. The principal reason for this difference
is not difficult to understand. Its root cause is the same physics 
responsible for the timeline in $R_{\rm h}=ct$ allowing large galaxies
to form by redshift $\sim 17$, as recently discovered by {\it JWST}.
Though the age of the Universe is about the same today in both models,
roughly equal to $1/H_0$, the time versus redshift relation was stretched
out by about a factor 2 at $z\gtrsim 6$ in $R_{\rm h}=ct$. The linear
growth of structure therefore extended for longer in this model 
compared to $\Lambda$CDM, allowing structure to grow to much larger
scales. Not only did large, well-formed galaxies appear at higher
redshifts in $R_{\rm h}=ct$, but the deviation from homogeneity due
to clustering was also enhanced by about an order of magnitude
in scale. 

\section{Discussion}\label{discussion}
At face value, the outcome with $R_{\rm h}=ct$ would appear to be much 
more consistent with the existence of very large structures (Table~\ref{tab1}) 
than the corresponding case in $\Lambda$CDM. It is also worth mentioning that
observational evidence of anomalously large structures is independently
provided by the measurement of the impact of large dark matter
fluctuations on the CMB via the integrated Sachs-Wolfe effect 
(ISW) \citep{Flender:2013}. The magnitude of the observed signal 
is more than $3\sigma$ larger than the theoretical $\Lambda$CDM
expectation, indicating that dark matter inhomogeneities on 
scales beyond $\sim 100\;{\rm Mpc}/h$ are larger than expected.

As noted earlier, however, there are several caveats to this 
simple-minded comparison. We may not have seen the largest 
structure yet. Though in principle all such features discovered 
thus far are consistent with the clustering deviation from 
homogeneity in $R_{\rm h}=ct$, we would need to revisit this 
interpretation should future, deeper surveys uncover yet larger 
structures. Then the question of whether FLRW is truly the correct 
metric to use in cosmology would become a pressing issue. 

The more serious caveat, though, is whether the discovery of very
large structures should be viewed as a violation of the {\it averaged}
transition to smoothness on smaller scales. At a very minimum, we
should ask whether the linear growth phase had sufficient time to produce
such large clustering with $\Lambda$CDM as the background cosmology.
It certainly did in the case of $R_{\rm h}=ct$. But what about the
standard model?

The answer may depend on the actual scale of the large structures.
For example, the Sloan Great Wall has been known for almost twenty
years \citep{Gott:2005}. This filamentary structure identified in
the Sloan Digital Sky Survey galaxy distribution extends over
$\sim 450$ Mpc and has been viewed as an extremely unlikely
occurrence in $\Lambda$CDM \citep{Sheth:2011}. Nevertheless, some 
$N$-body simulations show that structures as large as this do 
emerge in a $\Lambda$CDM background \citep{Park:2012}. Still, the 
probability for forming even larger structures, such as the Giant 
Arc \citep{Lopez:2022} and beyond, drops precipitously as the size 
increases.

An argument is sometimes made that structures such as the Huge-LQG
\citep{Clowes:2012} are so physically large that they could not
represent gravitationally bound systems in standard cosmology---this
is another way of saying that there would not have been sufficient
time for them to grow within the standard timeline. And given that 
clustering algorithms may find such extended features even in pure 
Poisson noise then means that, though they may be seen in the data, 
they do not actually represent real physical correlations 
\citep{Nadathur:2013}. This is certainly true, but it ignores 
the possibility that the background cosmology may simply not 
be $\Lambda$CDM. As noted earlier, the timeline beyond $z\sim 6$
is stretched by about a factor $2$ in $R_{\rm h}=ct$ and, as we
can see in Figures~(\ref{fig2}) and (\ref{fig3}c), the linear
growth phase in this model could adequately permit such large
structures to grow gravitationally by the redshift at which we
see them today.

The stark contrast seen between $\Lambda$CDM and
$R_{\rm h}=ct$ in Figures~(3b) and (3c) will likely also impact
the expected appearance of our past cosmological lightcone
\citep{Carfora:2021,Carfora:2022}. The issue here is how this
lightcone is modified as a function of proper distance from the
observer, given that the assumption of homogeneity in the FLRW
metric breaks down on small spatial scales. A formal definition
of the differences one may expect bears on several key observational
signatures, including the angular diameter distance and lensing 
distortions. 

A chief difficulty with the interpretation of
the FLRW spacetime geometry arises when past lightcone data
are gathered in our cosmological neighborhood, where the
matter distribution becomes highly anisotropic with a
high density contrast. On scales where this lack of
smoothness becomes predominant, the Einstein evolution of
the FLRW spacetime is, at best, an approximation, certainly
not an accurate representation of the actual dynamics.
\cite{Carfora:2022} characterize this in terms of the
distance-dependent level of tension one ought to observe
between the model predictions and the observations, such
that different models with different scales at which the
transition to homogeneity is expected should experience
measurably different levels of discordance arising from
the simplifying assumption of homogeneity everywhere. 
Thus, in addition to the expected variation of the smoothness
transition scale from one model to the next, it is reasonable
to anticipate a distance-dependent variation of the tension
with future high precision measurements that extends to larger
scales when the real Universe is more anisotropic than the 
model predicts.

\section{Conclusion}\label{conclusion}
More and more we are seeing growing tension between the formation
of structure predicted by the standard model and the actual
observations. This is manifested in the surprisingly rapid formation
of supermassive black holes, the too early appearance of galaxies,
and the incorrect mass distribution of dark matter halos at 
$z\gtrsim 4$ (see \citealt{Melia:2022e}, and references cited 
therein). The disparity between theory and the data has reached
a high point with {\it JWST}'s recent detection of large, well-formed 
galaxies at $z\sim 17$ 
\citep{Pontoppidan:2022,Finkelstein:2022,Treu:2022,Robertson:2022}. 

In all these cases, the timeline expected within the $R_{\rm h}=ct$
cosmology not only mitigates the tension, but appears to be fully
consistent with the evolution of all these systems
\citep{Melia:2014a,MeliaMcClintock:2015,Yennapureddy:2018a}. As an 
advanced version of $\Lambda$CDM, it is therefore sufficiently well
supported by the data for us to continue its development and testing. 
In this paper, we have taken the next step by examining the impact
of its extended timeline on the formation of the largest structures.
We have found that the transition to homogeneity in this cosmology
is expected to occur at a spatial scale $\sim 2200\;{\rm Mpc}/h$, 
comfortably beyond all of the large features discovered in the
cosmos thus far. 

There are some indications, however, that none of the existing
catalogs are yet large enough for us to have seen the largest
structure \citep{DeMarzo:2021}. If future, deeper surveys find
correlations in the matter distribution well beyond 
$\sim 2200\;{\rm Mpc}/h$, we may have to seriously rethink
the suitability of the FLRW metric for the description of our
cosmic spacetime.

\section*{Acknowledgments}
I am grateful to Giordano De Marzo for helpful 
comments that have led to an improvement in the presentation 
of this material.

\section*{DATA AVAILABILITY STATEMENT}
No new data were generated or analyzed in support of this research.

\bibliographystyle{mn2e.bst}
\bibliography{ms.bib}

\label{lastpage}
\end{document}